\documentclass[final, 5p, times,  twocolumn]{elsarticle}

\usepackage{listings}
\usepackage{xcolor} 
\usepackage{pgfplots}
\usepackage[most]{tcolorbox}
\usepackage{lipsum} 

\definecolor{codegreen}{rgb}{0,0.6,0}
\definecolor{codegray}{rgb}{0.5,0.5,0.5}
\definecolor{codepurple}{rgb}{0.58,0,0.82}
\definecolor{backcolour}{rgb}{0.95,0.95,0.92}

\lstdefinestyle{mystyle}{
 backgroundcolor=\color{backcolour},  commentstyle=\color{codegreen},
 keywordstyle=\color{magenta},
 numberstyle=\tiny\color{codegray},
 stringstyle=\color{codepurple},
 basicstyle=\ttfamily\footnotesize,
 breakatwhitespace=false,     
 breaklines=true,         
 captionpos=b,          
 keepspaces=true,         
 numbers=left,          
 numbersep=5pt,         
 showspaces=false,        
 showstringspaces=false,
 showtabs=false,         
 tabsize=2
}

\usepackage{graphicx}

\usepackage{algorithm2e}
\usepackage{multirow}
\usepackage{booktabs}
\usepackage{amsmath,amssymb,amsfonts}
\usepackage{algorithmic}
\usepackage{graphicx}
\usepackage{textcomp}
\usepackage{booktabs}
\usepackage{rotating}
\usepackage{caption}
\usepackage{multirow}
\usepackage[most]{tcolorbox}
\usepackage{url}%

\usepackage{amsmath}

\usepackage{comment}
\usepackage{float}
\usepackage{xcolor}
\usepackage{diagbox}
\usepackage{epstopdf}
\usepackage{caption}
\usepackage{subfig}

\usepackage{hyperref}
\usepackage{blindtext}
\usepackage{pifont}
\newcommand{\praioritize}{PrAIoritize }
\usepackage{xcolor}
\def\BibTeX{{\rm B\kern-.05em{\sc i\kern-.025em b}\kern-.08em
    T\kern-.1667em\lower.7ex\hbox{E}\kern-.125emX}}
    \pgfplotsset{compat=1.18} 
\begin{document}

\begin{frontmatter}

 \title{\praioritize: Automated Early Prediction and Prioritization of Vulnerabilities in Smart Contracts}

\author{Majd Soud\textsuperscript{1},
Grischa Liebel\textsuperscript{1},
and Mohammad Hamdaqa\textsuperscript{1,2}
}
\ead{majd18@ru.is, grischal@ru.is, mhamdaqa@ru.is}

\address{\textsuperscript{1}Department of Computer Science, Reykjavik University, Reykjavik, Iceland}
\address{\textsuperscript{2}Department of Computer and Software Engineering, Polytechnique Montreal, Montreal, Canada}
\begin{abstract}
\textbf{Context:} Smart contracts are prone to numerous security threats due to undisclosed vulnerabilities and code weaknesses. In Ethereum smart contracts, the challenges of timely addressing these code weaknesses highlight the critical need for automated early prediction and prioritization during the code review process. Efficient prioritization is crucial for smart contract security.\\
\textbf{Objective:} Toward this end, our research aims to provide an automated approach, \praioritize, for prioritizing and predicting critical code weaknesses in Ethereum smart contracts during the code review process.\\
\textbf{Method:} To do so, we collected smart contract code reviews sourced from Open Source Software (OSS) on GitHub and the Common Vulnerabilities and Exposures (CVE) database. Subsequently, we developed PrAIoritize, an innovative automated prioritization approach. \praioritize integrates advanced Large Language Models (LLMs) with sophisticated natural language processing (NLP) techniques. \praioritize automates code review labeling by employing a domain-specific lexicon of smart contract weaknesses and their impacts. Following this, feature engineering is conducted for code reviews, and a pre-trained DistilBERT model is utilized for priority classification. Finally, the model is trained and evaluated using code reviews of smart contracts.\\
\textbf{Results:} Our evaluation demonstrates significant improvement over state-of-the-art baselines and commonly used pre-trained models (e.g. T5) for similar classification tasks, with 4.82\%-27.94\% increase in F-measure, precision, and recall.\\
\textbf{Conclusion:} This paper introduces \praioritize, an automated approach that effectively prioritizes smart contract code weaknesses during the review process. By leveraging \praioritize, practitioners can efficiently prioritize smart contract code weaknesses, addressing critical code weaknesses promptly and reducing the time and effort required for manual triage.

\end{abstract}


\end{frontmatter}

\section{Introduction}

Smart contracts are Turing-complete 
programs operating on the blockchain to manage the flow of assets among the contractual parties~\cite{luu2016making}. Smart contracts shifted blockchain technology such as Ethereum from primarily storing and transferring cryptocurrencies (e.g., Bitcoin) to enabling a wide range of transactions and decentralized applications (DApps) in various fields~\cite{chen2021maintenance}. Smart contracts are typically written in high-level programming languages such as Solidity~\cite{wood2014ethereum}.
 

Inheriting immutable, self-executing, and decentralized attributes from the underlying blockchain technology, smart contracts cannot be modified once deployed to the blockchain, and their execution is entirely contingent on their unchangeable code~\cite{wood2014ethereum}.
In Ethereum, modifying smart contracts to fix vulnerabilities can be costly and complex, involving deploying new versions of contracts and upgrading them~\footnote{https://docs.openzeppelin.com/learn/upgrading-smart-contracts}.
These inherent characteristics ensure the reliability of smart contracts, and set them apart from conventional software.

The distinguishing features of smart contracts make them vulnerable to security threats.
%
For instance, Ethereum's unique gas system and smart contract immutability exacerbate existing security threats~\cite{bosu2019understanding}, e.g., often resulting in significant financial losses in case of an attack.
Notable incidents such as the DAO attack, leading to the hack of about 50 million dollars, highlight the substantial risks associated with smart contract code weaknesses~\cite{mehar2019understanding}. 

Challenges in smart contract maintenance extend beyond inherent code weaknesses. 
Automated smart contract security tools can help mitigating a significant number of attacks \cite{chaliasos2024smart}, but the large number of audits and reviews generated by these tools presents a significant challenge, particularly for auditors tasked with manual triage. 
Hence, there is a need for improved automation in triaging and prioritizing code weaknesses, towards building tools that are useful for practitioners \cite{chaliasos2024smart}. 
Finally, the novelty of smart contract and blockchain technology can result in a substantial amount of weaknesses that are unknown, thus increasing the urgency of prioritizing weaknesses and addressing critical ones.

Therefore, this paper aims to answer the following research question:
\begin{itemize}
\item[RQ1:] \textit{To what extent can smart contract code weaknesses be successfully prioritized during code review processes? 
}
\end{itemize}
To answer this question, we present an automated approach to prioritize smart contract code weaknesses, called \praioritize.
We start by quantifying the occurrence of zero-day attacks, attacks that are unknown prior to their disclosure, in Ethereum smart contract code by studying the timing of exploit disclosure in records of the Common Vulnerabilities and Exposures (CVE) database, thus providing a stronger motivation for our work and future work on automated triage in smart contract code.
Then, we leverage a combination of Large Language Models (LLMs) and natural language processing techniques (NLP) to label unlabeled code weaknesses automatically and prioritize them.
Smart contracts, being sequential code with dependencies among code statements, are similar to natural language text. Moreover, they often follow specific templates and standards, leading to recognizable patterns and repetitive codes and structures~\cite{clack2016smart}. Therefore, LLMs and other NLP-based approaches have significant potential in automating the detection of code weaknesses in smart contracts and comment generation~\cite{hu2023large, zhao2024automatic}. 
We constructed a smart contract code weakness lexicon that includes various code weaknesses and corresponding impacts. Subsequently, we utilized this lexicon to automatically label unlabeled code reviews. We then applied the DistilBERT~\cite{sanh2019distilbert}, to automatically prioritize smart contract code weaknesses.
Moreover, we have made the models, scripts and data used in this study openly accessible to encourage research replication and to facilitate further investigations by other researchers in the field.
Our evaluation results show that \praioritize  outperforms two state-of-the-art baselines and four popular and well-known pre-trained models for text classification, with 4.82\%-27.94\%  higher F-measure, 2.35\%-27.94\%   precision, and 3.57\%-27.94\%  higher recall.

\textbf{Paper Organization.} We introduce the motivation in Section~\ref{sec:motivation}. In Section~\ref{sec:review}, we present the terminology and related background. Our methodology is detailed in Section~\ref{sec:framework}. Our experimental evaluation is presented in Section~\ref{sec:results}. Section~\ref{sec:empirical} presents the results and findings from our empirical analyses, followed by the discussion and implications in Section~\ref{sec:implications}. Related work is discussed in Section~\ref{sec:related_work}. Potential threats to validity are examined in Section~\ref{sec:threats}. Finally, the paper concludes in Section~\ref{sec:conclusion}.

\section{Motivation}
\label{sec:motivation}
In this section, we offer real-world examples to highlight the necessity of predicting and prioritizing smart contract code weaknesses in code reviews at an early stage. In real-world scenarios, developers face significant challenges when it comes to prioritizing code reviews. This difficulty arises from the need to manually examine numerous code reviews submitted by developers~\cite{gousios2016work}. Failure to prioritize these reviews effectively can result in wasted time and effort for developers~\cite{fan2018early}. For instance, reviewers may devote significant time meticulously analyzing or addressing a code review, only to discover that it is of low priority or remains unused within the smart contract code. It is a common practice to categorize code reviews with code weaknesses into different priority levels, such as critical, high, medium, or low. These levels are assigned by developers based on various factors, including the type and severity of the code weaknesses addressed, their impact, and other relevant considerations~\cite{tian2013drone}. We elaborate on these priority levels for smart contract code reviews in Section~\ref{sec:framework} of our paper.

\begin{figure*}[ht]
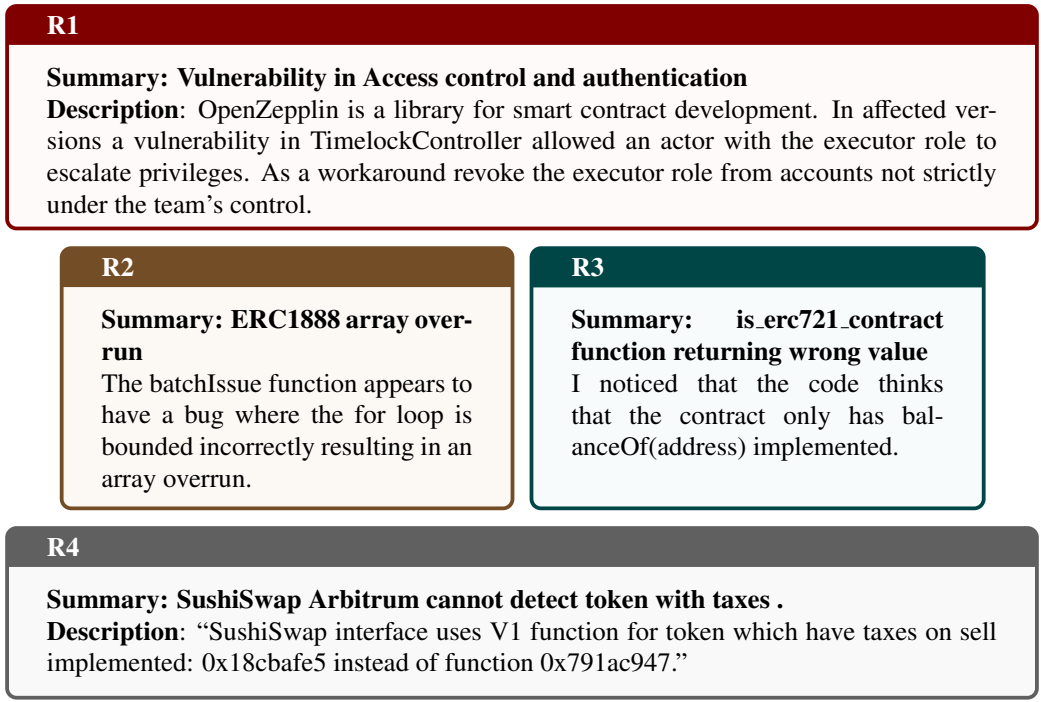

  \centering
  \begin{tcolorbox}[title = R1,colback=red!2!white,colframe=red!50!black,fonttitle=\bfseries,height=3cm, width = 13.6cm]
\textbf{Summary: Vulnerability in Access control and authentication
}\\
\textbf{Description}: 
OpenZepplin is a library for smart contract development. In affected versions a vulnerability in TimelockController allowed an actor with the executor role to escalate privileges. As a workaround revoke the executor role from accounts not strictly under the team's control.
\end{tcolorbox}

\begin{tcbraster}[raster equal height,raster valign=top,raster columns=3, raster rows=1,
colframe=brown!50!black,colback=brown!20!black,colbacktitle=brown!60!black]

 \centering
\begin{tcolorbox}[title =R2,colback=	brown!5!white,colframe=	brown!60!black,fonttitle=\bfseries,height=3.5cm, width = 7cm]

\textbf{Summary: ERC1888 array overrun}\\
The batchIssue function appears to have a bug where the for loop is bounded incorrectly resulting in an array overrun. 
 
 \end{tcolorbox}
  \centering
\begin{tcolorbox}[boxsep=0pt,boxrule=0pt,colback=white,colframe=white,enhanced jigsaw,left=0mm,right=0mm,top=0mm,bottom=0mm]
\begin{tcolorbox}[title = R3,colback=teal!3!white,colframe=teal!55!black,fonttitle=\bfseries,height=3.5cm, width = 6cm]

\textbf{Summary: is\_erc721\_contract function returning wrong value}
\\
 I noticed that the code thinks that the contract only has balanceOf(address) implemented.
\end{tcolorbox}

\end{tcolorbox}

\end{tcbraster}
\begin{tcolorbox}[title = R4,colback=gray!5!white,colframe=gray!75!black,fonttitle=\bfseries,height=2.30cm, width = 13.6cm]
\textbf{Summary: SushiSwap Arbitrum cannot detect token with taxes .
}\\
\textbf{Description}: ``SushiSwap interface uses V1 function for token which have taxes on sell implemented: 0x18cbafe5 instead of function 0x791ac947.''
\\

\end{tcolorbox}
\caption{Illustrated Examples: Real-World Smart Contract Code Reviews}
\label{fig:exampl}
\end{figure*}


 In Figure~\ref{fig:exampl}, we illustrate the four priority levels using examples of four real-world code reviews. In the first code review (R1)~\cite{report1}, a vulnerability is identified in the TimelockController. This vulnerability enabled an actor with the executor role to gain immediate control of the timelock by resetting the delay to 0 and escalating privileges. Consequently, the actor could obtain unrestricted access to assets stored in the contract. Instances with the executor role set to ``open'' posed a particular risk, as they allowed any user to exploit the executor role, thereby leaving the timelock vulnerable to takeover by potential attackers. If such a vulnerability were present in a high-value contract, such as the Axie Infinity\footnote{Contract address: 0xbb0e17ef65f82ab018d8edd776e8dd940327b28b} contract with a net value of \$921 million \footnote{https://www.coingecko.com/en/coins/axie-infinity}, it could result in significant financial losses, client loss, and damage to reputation. Therefore, it is a critical priority code review, and it is crucial to fix the vulnerability to prevent attacks. Moreover, this vulnerability is reported and disclosed in the CVE record.


 The second code review (R2)~\cite{report2} demonstrates a code weakness in a function in the Registry contract. It is caused by an incorrectly bounded loop, which eventually may result in an array overrun or extra gas~\footnote{The unit for measuring the computational effort required to execute a transaction on Ethereum.} consumption. While this specific weakness does not immediately expose the contract to exploitation or jeopardize control over its functionality (i.e., critical), it nonetheless demands high attention due to its impact on gas consumption within the contract. Excessive gas consumption in smart contracts not only results in the loss of Ether~\footnote{The native cryptocurrency for the Ethereum blockchain.} but also leads to poor contract performance, potentially affecting user experience. On its own, the code weakness may not directly provide attackers with a means to compromise the contract's integrity or gain unauthorized access. Instead, it represents a potential avenue for exploitation when coupled with other vulnerabilities or when attackers leverage specific circumstances to exploit the weakness. Although deemed less critical than the vulnerability discussed in the first example, this weakness remains a high-priority weakness due to its significant potential impact on both the performance and security of the contract.

 

In the third review~\cite{report3}, a function is used to determine whether a contract adheres to ERC721 standard contract~\footnote{A widely adopted smart contract standard for representing ownership of unique non-fungible tokens~\cite{ERC721}.} or not. The code weakness is prompted when the function is invoked within the contract bytecode.
If the code weakness is triggered, we consider it to be of medium priority as it can have notable implications on the contract interactions, especially those involving dependent contracts that rely on this information. Specifically, if triggered, the weakness may cause dependent contracts to incorrectly perceive the contract as non-compliant with ERC721 standards, potentially leading to cascading functional issues and possibly financial losses during interactions. 
This weakness, with its localized impact and lack of immediate exploitability, poses less imminent threats compared to previous examples that present more immediate risks, thus warranting higher priorities.

Finally, the last review (P4)~\cite{report4} identifies a code weakness with the SushiSwap contract, one of the biggest decentralized exchanges (DEX), where the interface shows the wrong function for tokens with taxes, resulting in an error displayed in the console. The code weakness does not affect contract functionality or associated dependencies. Therefore, we consider this code review as a low-priority. Compared to the aforementioned code reviews, this code weakness is less critical and does not pose financial losses. 

The definitions of the proposed different priority levels are provided in Section~\ref{sec:framework}, while the limitations of this classification are listed in Section~\ref{sec:threats}.

\section{Background and Terminology}\label{sec:review}
In this section, we provide essential background information and define key terms relevant to our research. We also provide background on smart contract security and zero-day attacks, highlighting the urgency of addressing vulnerabilities in smart contracts. We outline the primary objectives of code reviews and introduce pre-trained models utilized for automating the prioritization and prediction of code weaknesses in the existing literature.

\subsection{Definitions}
\label{sec:def}

\begin{itemize}
    \item \textbf{A vulnerability}: ``A weakness in the computational logic (e.g., code) found in software and hardware components that, when exploited, leads to adverse effects on confidentiality, integrity, or availability. Addressing such weaknesses usually requires modifications to the codebase, but may also entail alterations to specifications or even the deprecation of certain specifications (e.g., eliminating affected protocols or functionalities entirely)''~\cite{nvdVul}.

\item \textbf{Weakness:}
     ``a software error or mistake in contract code that in the right conditions can by itself or coupled with other weaknesses lead to a vulnerability''~\cite{EIP1470}). 
    
\item \textbf{Priority}: Refers to the level of urgency assigned to a software code weakness, indicating how quickly the code weakness needs to be fixed and removed. Priority levels, such as high, low, medium, and critical, are commonly used to represent the urgency of addressing the weakness. However, the specific definitions of these priority levels may vary based on the field. The priority level is typically determined from the perspective of the software developers and is based on several factors, such as weakness severity, potential impact on system functionality, and business criticality~\cite{tian2013drone}.

\item \textbf{Exploit:} A piece of software, or a sequence of commands that takes advantage of a code weakness, glitch, or vulnerability in order to cause unintended or unanticipated behavior to occur in computer software or hardware. It is commonly used to gain unauthorized access to a system, execute arbitrary code, or perform other malicious activities. 

\item \textbf{Third Party Advisory:} A notification or report issued by an external entity (e.g., a security researcher, a cybersecurity organization) regarding a security vulnerability or weakness found in a software product, system, or service. These advisories typically provide details about the vulnerability, its potential impact, and guidance on how to mitigate the risk.

\item \textbf{Vendor Advisory:} A formal communication issued by the vendor or developer of a software product, system, or service to alert users about security vulnerabilities, bugs, or weaknesses identified in their products.

\end{itemize}

\begin{table*}[htbp]
  \centering
  \caption{Explanation of the collected CVE Attributes}
  \label{tab:cve_attributes}
  \begin{tabular}{@{}ll@{}}
    \toprule
    \textbf{Attribute}              & \textbf{Explanation}                                                                                  \\ \midrule
    CVE ID                           & Unique identifier assigned to a CVE entry.                                                            \\
    Publication Date                 & Date when the CVE entry was published.                                                               \\
    Last Modified Date               & Date when the CVE entry was last modified.                                                           \\
    CVE Description                  & Description of the vulnerability or weakness described by the CVE entry.                               \\
    Severity                         & Severity level assigned to the CVE entry, indicating the potential impact of the vulnerability.      \\
    CVSS2/CVSS3 Access Complexity   & Complexity of access required to exploit the vulnerability.                                            \\
    CVSS2/CVSS3 Authentication      & Authentication required to exploit the vulnerability.                                                 \\
    CVSS2/CVSS3 Confidentiality     & Impact on confidentiality if the vulnerability is exploited.                                           \\
    CVSS3 Attack Vector              & Vector that describes how the vulnerability can be exploited.                                           \\
    CVSS3 Attack Complexity          & Complexity of the attack required to exploit the vulnerability.                                         \\
    CVSS3 Integrity Impact           & Impact on integrity if the vulnerability is exploited.                                                  \\
    GitHub Link                      & Link to the GitHub repository or issue related to the CVE entry.                                         \\
    Exploit Date                     & Date when the vulnerability was exploited, if applicable.                                               \\
    Third Party Advisory Date        & Date of a third-party advisory related to the CVE entry, if available.                                   \\
    Patch Date                       & Date when a patch or fix for the vulnerability was released, if available.                                             \\ \bottomrule
  \end{tabular}
\end{table*}

\subsection{Smart Contract Security}
 Smart contracts, integral to blockchain technology, facilitate decentralized execution of agreements and transactions without intermediaries. They have transformed various sectors by enhancing transparency, trust, and efficiency. Solidity, a programming language tailored for smart contracts, empowers developers to define their logic on platforms such as Ethereum. Despite the benefits that smart contracts provide, they are vulnerable to various security threats \cite{perez2019smart}. Furthermore, there have been numerous instances of smart contract security breaches in recent years \cite{atzei2017survey}. The first attack occurred in 2016, when an attack on DAO contracts resulted in the loss of over 3.6 million Ethers due to a re-entrancy vulnerability. In 2020, the security research team at CertiK discovered several vulnerabilities in the smart contract for the SushiSwap project \cite{berg2022empirical} that allow the owner of the contract to perform unauthorized actions. Due to increasing attacks, smart contract security and trustworthiness have gained significant attention from scholars, resulting in numerous studies on smart contract vulnerabilities.

\subsection{Common Vulnerability and Exposure (CVE) and National Vulnerability Database (NVD)}
CVE (Common Vulnerabilities and Exposures)~\footnote{https://cve.mitre.org/} is a catalog of publicly disclosed vulnerabilities and exposures managed by MITRE. It synchronizes with the NVD (National Vulnerability Database)~\footnote{https://nvd.nist.gov/vuln}, ensuring continuous alignment between the two. The NVD serves as a comprehensive repository containing information on all publicly known software vulnerabilities. It offers detailed insights into CVE-listed vulnerabilities, including exploit date, patch availability, and various search functionalities. Table~\ref{tab:cve_attributes} presents the attributes gathered from NVD for each CVE. It includes impact metrics such as the Common Vulnerability Scoring System (CVSS), vulnerability types categorized under the Common Weakness Enumeration (CWE), and other relevant metadata.
Each CVE entry is assigned a unique identifier delineating the affected software product, sub-products, and different versions.

\subsection{GitHub Code Reviews}
A code review, also known as a peer review, is a systematic examination of software source code by one or more individuals other than the author(s) with the aim of identifying defects, improving quality, and transferring knowledge. It is a fundamental practice in software development aimed at ensuring the reliability, maintainability, and security of the codebase~\cite{bacchelli2013expectations}. Code reviews typically involve a team of developers collaboratively inspecting a piece of code line by line to identify code weaknesses such as vulnerabilites, defects, logical errors, performance bottlenecks, and violations of coding standards or best practices.

Code reviews offer several benefits, including early detection and resolution of code weaknesses, knowledge sharing among team members, and improvement of overall code quality.  

A typical code review on GitHub encompasses various components, including a description of code weaknesses, a summary, auditor comments, code weakness details, dependency information, priority, severity, assignee, and proposed fixes. However, in the context of smart contracts, not all of these elements may be present. In our analysis, we focus on essential elements available in GitHub, particularly the Description field. These fields contain valuable information crucial for training our model and predicting the priority of code weaknesses.

 \subsection{Zero-day attacks} 
\begin{figure}[h!]
    \centering
        \includegraphics[width=0.4\textwidth]{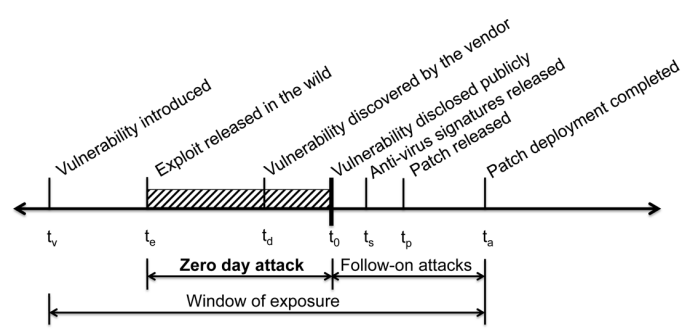}
    \caption{Timeline of Zero-day Attacks~\cite{bilge2012before}.}
    \label{fig:ZDAt}
        
\end{figure}
Zero-day attacks represent a class of cyber threats characterized by the exploitation of software or system vulnerabilities before vendors can develop and distribute patches or fixes. These attacks exploit vulnerabilities that are unknown to the vendor or for which no mitigating measures are available at the time of the attack. Consequently, zero-day attacks present a formidable challenge to organizations and individuals due to their ability to bypass existing security mechanisms without prior warning, potentially leading to data breaches, system compromises, and other security incidents with significant repercussions.

A zero-day attack timeline,  as shown in Figure~\ref{fig:ZDAt},  typically begins with the introduction of a vulnerability into software (\textit{time= tv}). Followed by the release of an exploit in the wild by malicious actors (\textit{time= te}). Once the vendor becomes aware of the vulnerability, they assess its impact and begin working on a patch(\textit{time = td}). The vulnerability is then publicly disclosed, assigned a CVE identifier (\textit{time = t0}). Then, anti-virus signatures are released to detect ongoing attacks (\textit{time = ts}). Finally, the vendor releases a patch (\textit{time = tp}), and once it is deployed on all vulnerable hosts, the vulnerability ceases to have an impact (\textit{time = ta}). A zero-day attack occurs when the vulnerability is exploited before it is publicly disclosed, i.e., \textit{t0~\textgreater~te}, highlighting the importance of timely patching. One of our goals in this paper is to measure the prevalence of zero-day attacks and assess the effectiveness of patching code weaknesses on zero-day vulnerabilities before and after disclosure~\cite{bilge2012before}. 

It is important to note that in smart contracts, the concept of releasing viruses is not applicable. Instead, when vulnerabilities are identified, fixes are released to address the code weakness and enhance security.

\subsection{Related Models} \label{sec:pretrained_models}
Pre-trained language models have achieved remarkable success in various software engineering tasks~\cite{yang2022survey,yang2015deep}, such as code summarization and code weakness localization. This section briefly describes state-of-the-art pre-trained models that have been used to perform classification tasks similar to our task and achieved high performance.

Bidirectional Encoder Representations from Transformers (BERT) is a pre-trained language model proposed by Google~\cite{devlin2018bert} that has achieved state-of-the-art performance on different software engineering tasks (e.g.~\cite{ciborowska2022fast}). BERT can learn dynamic context word vectors and capture textual semantic features. 

DistilBERT~\cite{sanh2019distilbert} is designed to be more memory efficient and faster to train than BERT. 

T5 (Text-To-Text Transfer Transformer) is a language model developed by Google AI Language. It belongs to the Transformer architecture family~\cite{vaswani2017attention}. T5 stands frames all tasks as text-to-text transformations~\cite{raffel2020exploring}. This means that T5 is trained to input a text prompt and generate the corresponding output text. Its applications include various tasks such as translation, summarization, question answering, and text generation, as text transformation problems.


Bidirectional Long Short-Term Memory (BiLSTM) is a popular neural network architecture widely used for natural language processing tasks, including text classification~\cite{liu2019bidirectional}. As BiLSTM processes the input sequence in both directions, it can capture past and future contexts. 

Recurrent Neural Networks (RNNs) are commonly used for text classification tasks~\cite{lipton2015critical}. RNNs are designed to effectively capture input text sequential dependencies. 

Bidirectional Long Short-Term Memory (BiLSTM) is a popular neural network architecture widely used for natural language processing tasks, including text classification~\cite{liu2019bidirectional}. As BiLSTM processes the input sequence in both directions, it can capture past and future contexts.

 \section{Methodology}
\label{sec:framework}
In this section, we outline the approach we took to investigate zero-day attacks in smart contracts. Next, we detail the different stages of our automated prioritization method, \praioritize.

\begin{figure*}[ht]
\centering
\includegraphics[width=1.0\textwidth]{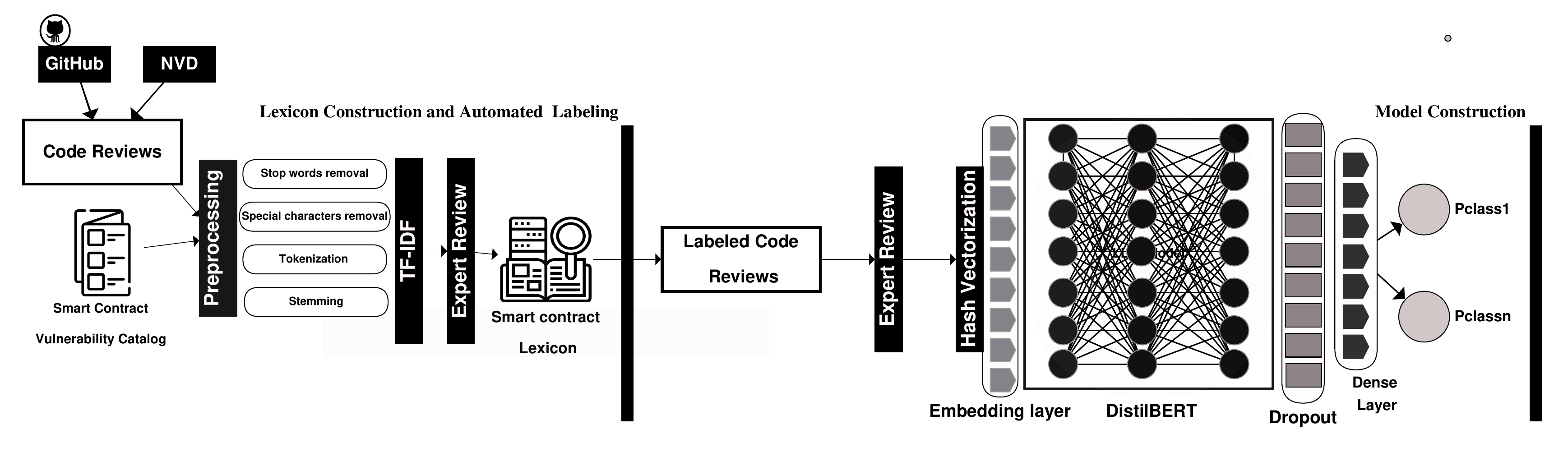}
\caption{\praioritize Approach Overview}
\label{fig:overview}
\end{figure*}

\subsection{Definition of Priority Levels for Smart Contract Weaknesses}
In our study, we define the priority levels for smart contract code weaknesses based on an analysis we conducted in a previous research and our own findings~\cite{soud2022fly}. Specifically, we conducted an analysis of code weaknesses in smart contracts, drawing insights from our examination of various sources, including Stack Overflow, Github, CVE, and other relevant literature. Through this process, we identified prevalent code weaknesses and corresponding impacts in smart contract code. Building on these insights, we define the four priority levels in smart contracts, taking into account several factors. First and foremost is \textbf{exploitability}, assessing the extent to which attackers could capitalize on a weakness to compromise the contract integrity. \textbf{Impact} also plays a crucial role, analyzing the extent of potential harm, whether it is financial losses or performance issues. Additionally, the \textbf{locality} of effects is evaluated to determine whether a weakness's impact is confined to specific components or has broader implications. The \textbf{immediacy of threat} is another important consideration, considering the urgency of addressing the weakness to mitigate potential risks promptly. Finally, \textbf{dependencies} are examined to understand how reliant other libraries or contracts are on the vulnerable component and the potential effects of exploitation. By considering these factors, we define four levels of code weakness priority as follows:
      \begin{itemize}
      \item \textbf{Critical priority}: The code weakness present in the smart contract code. It poses an immediate risk to the contract's integrity, security, and functionality, potentially resulting in the loss of access to the entire contract.  These weaknesses are highly exploitable by attackers and result in significant financial losses such as access control weaknesses. Critical priority weaknesses demand urgent attention and mitigation due to their high exploitability and severe impact on the contract's operation and security.

      \item \textbf{High priority}: The code weakness in the smart contract causes significant undesirable outcomes. In contrast to a critical weakness, it cannot be exploited directly by attackers. High priority weaknesses may be coupled with existing vulnerabilities, leading to cascading failures or performance degradation. 
      
      \item \textbf{Medium priority}: The code weakness is within the contract implementation and may result in unintended logic. While not directly exploitable by external attackers, these weaknesses can affect dependent contracts and may lead to incorrect functionality when interacting with other decentralized applications (DApps). Such weaknesses have the potential to disrupt the seamless operation of interconnected DApps and impact the overall interoperability of the smart contract ecosystem.
      
      \item \textbf{Low priority}: The code weakness does not affect the contract's functionality or any associated environment calls outside the contract. These code weaknesses may be related to unused variables in the contract code, errors in the contract's interface, documentation, or other non-essential components.
\end{itemize}

\subsection{Overall Approach}
Our approach starts with the data collection phase, where we collect CVEs from the NVD  database and code reviews from GitHub related to Ethereum smart contracts. We then conduct an investigation into zero-day attacks by quantitatively analyzing CVE records, including exploit dates, third-party advisory dates, vendor advisories, CVE patches, and common weaknesses (CWE). Motivated by the findings from our quantitative analysis, we propose \praioritize. \praioritize leverages LLMs and NLP to predict the priority levels of smart contract weaknesses. As shown in Figure~\ref{fig:overview}, \praioritize consists of three main phases: (1) lexicon construction and automated labeling phase, (2) classification model construction, and (3) training phase.

The first phase of our methodology involves collecting well-known smart contract code weaknesses and vulnerabilities in a smart contract vulnerability catalog, then using the vulnerability catalog along with the code reviews collected from GitHub and CVE to construct a lexicon of code weaknesses associated with impacts in smart contracts, as well as their respective priority levels. Followed by automatically generating priority labels for unlabeled code reviews using the constructed lexicon.
In the model construction, \praioritize utilizes feature engineering to capture textual and semantic factors that may impact the priority level of a code weakness. These features are then passed through a classification model to create a model capable of classifying a code review with an unknown priority level. Finally, in the prediction phase, \praioritize leverages the DistilBERT model to predict the priority levels of a given set of code reviews. The features of each code review are extracted, including tokenized text, embeddings, semantics, and other relevant characteristics, and are utilized by the model to analyze and predict the priority level. We further detail our approach in the following sections

\subsection{Dataset Collection}
To answer the proposed research question, we utilized AutoMESC~\cite{soud2023automesc} to extract data from two well-known publicly accessible sources; CVE from NVD and GitHub. These sources are commonly used for reporting vulnerabilities in Ethereum smart contracts and other software systems. Table~\ref{tab:dataset} describes the data sources and the final selected number of reviews per source after pre-processing. 

\textbf{Data-source 1--- CVE.} We utilized AutoMESC to collect and extract CVEs associated with smart contracts between 2018 and 2023 from the NVD database. The collected data include various attributes and metadata, such as CVE ID, publication date, last modified date, CVE description, severity, CVSS2 access complexity, CVSS2 authentication, CVSS2 confidentiality impact, CVSS3 attack vector, CVSS3 attack complexity, CVSS3 integrity impact, GitHub link, exploit date, third-party advisory date, and patch date. Table~\ref{tab:cve_attributes} identifies the collected attributes per CVE as described in the NVD website.

\textbf{Data-source 2--- GitHub.} We opted to utilize GitHub as our secondary data source for this study due to its status as the most widely used social coding platform, as highlighted in previous research~\cite{cosentino2017systematic}. Additionally, numerous studies exploring Ethereum smart contracts and related analysis tools have relied on data sourced from GitHub's open-source projects~\cite{durieux2020empirical}. Our data collection concentrated on identifying code weaknesses across various open-source projects featuring Ethereum smart contracts written in Solidity\footnote{All the smart contracts and the issues are listed in our artifact}.

We initially collected relevant code reviews corresponding to the collected CVEs from Data-source1 utilizing the provided Github Link for each CVE. Next, we employed targeted keywords such as ``smart contract", ``Solidity", and ``Ethereum". At the same time, we conducted searches using terms such as ``issue", ``defect", ``weakness", ``problem", and ``vulnerability" to broaden our collection of issues related to Ethereum smart contracts. Additionally, we included the Ethereum official GitHub repository in our analysis.

Then, we excluded code reviews with open issues in smart contracts, concentrating our efforts on thoroughly code reviews and confirmed weaknesses as evident in the closed issue reviews. This decision was made to ensure a focused examination of verified weaknesses and maintain the integrity of our research findings. Our focus was specifically on newly published code reviews within the four years, up until February 2023, to capture the most recent and relevant information.
For noise reduction, code reviews with fewer than 5 words were excluded. The detailed number of code reviews per priority level is also listed in Table~\ref{tab:dataset}. 
\begin{table}[ht!]
\centering
\small
\caption{Summary of the sampled smart contract reviews with code weaknesses in the selected data sources}
\label{table:datasetSummary}
\begin{tabular}{p{.2\textwidth} p{.060\textwidth} p{.060\textwidth}p{.06\textwidth}} 
 \toprule
\textbf{Data Source} & \textbf{GitHub} & \textbf{NVD } & \textbf{Total}\\ [0.5ex]
 \toprule
 \# of collected data& 2100 &442  & 2542\\

 \# of data records after preprocessing& 1000 & 440& 1440\\
\toprule
\end{tabular}
\label{tab:dataset}
\end{table}

 Furthermore, the dataset is automatically and randomly divided into training and testing sets, with an 80/20 ratio, respectively.

\subsection{Investigating zero-day attacks in smart contracts}
To identify zero-day attacks, we leveraged the attributes collected from \textit{Data-source 1 }for each CVE. We focused on extracting key dates, including the exploit date, the third-party advisory date, the vendor advisory date, and the patch date whenever available. Moreover, we conducted an in-depth analysis of the temporal relationships among these dates. Precisely, our analysis involved examining the proximity between the exploit date and the other dates (i.e., the third-party advisory, vendor advisory, and patch dates) following the timeline discussed in the background section~\ref{sec:review}. This analysis enabled us to pinpoint the vulnerabilities that indicate zero-day attack scenarios.

The analysis was done automatically, by retrieving input data from the specified source and extracting key dates. If an exploit date is absent, but an advisory date exists, indicating no attack, the analysis proceeds with the next CVE record. However, if both exploit and advisory dates are identified, the temporal difference is analyzed. Instances where the exploit date precedes or coincides with the advisory date signify zero-day attacks. Furthermore, our automatic analysis detects instances where fixes were implemented before attacks occurred, indicating preemptive patching of vulnerabilities. Ultimately, statistics on zero-day attacks within the dataset are calculated, offering insights into the prevalence of such vulnerabilities in smart contracts.

\subsection{Data Cleaning and Preprocessing}
\label{sec:dcleaning}
In this phase, each code review is preprocessed and cleaned to ensure that it does not have any data quality issues that may negatively affect the classification and prediction results. Code reviews consist of textual descriptions and code snippets, and it is crucial to process the textual description with care, as even a small error can lead to the exclusion of useful information.

\begin{table}[ht!]
\centering
\small
\caption{Statistics of Collected Code Weaknesses per Priority Level}
\begin{tabular}{@{}lllll@{}}
\toprule
\multicolumn{5}{c}{\textbf{ Code weaknesses per Priority Level}}                                  \\ \midrule
 & \textbf{Low}            & \textbf{Medium}            & \textbf{High}         & \textbf{Critical}         \\ \midrule
\textbf{Total} &  192                     & 179                        & 234                   & 395                       \\ \bottomrule
\end{tabular}
\label{tab:datasetperlevel}
\end{table}

We first extracted the code weakness description from each code review. Then, we removed all special characters and punctuation characters from each code review. These elements have no significant effect on the text mining process, and removing them from the text data ensures optimal performance of the text mining algorithms~\cite{bermingham2015application}.

We then manually remove the URLs, code snippets, configurations, and transaction logs to ensure that the model can better comprehend the textual description. This is done to minimize the potential for the model to be confused by extraneous information and to better focus on the main content of the code review. By simplifying the code review in this way, we aim to improve the model's ability to accurately predict the priority level of the code weaknesses. We outline the primary techniques that we employ to preprocess in the following paragraphs.

\textbf{Tokenization:} we utilized tokenization to break up a continuous stream of text into individual units called tokens. In the case of preprocessing the code reviews, the textual description is tokenized, meaning it is broken down into individual tokens, each of which refers to a word in the description. The extracted words are separated by delimiters, which in our case are white spaces.

\textbf{Stop Words Removal:} These are commonly used words that may not carry much meaning in the context of the code reviews, such as pronouns. We utilized a pre-existing list~\cite{loper2002nltk} of stop words to identify and remove them from the extracted corpus of code reviews.

\textbf{Stemming:} In the English language, words have multiple forms, making analysis challenging. We used stemming to resolve this problem by converting words to their root form. Stemming involves reducing words to their base form, (i.e., ``attack" is the base form of ``attacking'', ``attacked'', and ``attack''), which aids in analyzing the meaning of the text.  

\subsection{Automated Labeling}
Due to the increasing complexity of smart contracts, the need for an automated labeling approach that can efficiently and accurately prioritize code reviews has become more urgent. Automated labeling has been used in various Natural Language Processing (NLP) tasks in the literature~\cite{liu2010sentiment}, such as sentiment analysis~\cite{liu2010sentiment,kim2004determining} and text summarization~\cite{galappaththi2021automatically}.


In this study, we propose lexicon-based automatic labeling to assign priority labels  to unlabeled code reviews.

\textbf{Lexicon Construction.} One of the essential steps in our approach is constructing a lexicon of smart contract vulnerability-related keywords along with their priority levels. 
The proposed lexicon consists of keywords, and each keyword has a priority level. 
Lexicon construction process consists of several steps as shown in Figure~\ref{fig:overview} and Algorithm~\ref{algo1}. First, a vulnerability catalog is assembled by collecting well-known smart contract vulnerabilities and code weaknesses from the literature~\cite{soud2022fly} and from DASP\footnote{https://dasp.co/} which is a popular source of the top 10 critical security risks in smart contracts to include them in the lexicon. After that, the vulnerability catalog and the corpus of code reviews from the NVD and GitHub are preprocessed and concatenated using the techniques discussed in section~\ref{sec:dcleaning} as shown in Algorithm~\ref{algo1}-step 1. Then, index term extraction and term-weighting are applied using the term frequency-inverse document frequency (TF-IDF) weighting algorithm~\cite{debole2003supervised} and Bag of Word (BoW)~\cite{salton1988term} to extract keywords from the textual descriptions of the preprocessed corpus of code reviews and vulnerability catalog (i.e., Algorithm~\ref{algo1}-step 2). This step includes assigning each term in the textual code review with a numerical score using the TF-IDF weighting scheme that ranks the terms based on both the frequency of the word in each code review (term frequency) and the rarity of the word in the entire corpus (inverse document frequency). In addition to representing the textual descriptions of the code reviews as a bag (i.e., multiset) of its words while considering the frequency of each word in the review (i.e., Algorithm~\ref{algo1}-step 3). The resulting keywords are sorted based on their term frequency. Then, an expert—specifically, a doctoral student specializing in software engineering with a focus on smart contract security and software security~\footnote{The first author of this paper.}—assigned priority levels to the top 250 keywords. This assignment takes into account their respective impacts as identified in the code reviews. The lexicon is constructed based on the resulting keywords with their priority levels. The lexicon can enable efficient automatic labeling of smart contract reviews. After employing a randomization script, the expert meticulously reviewed the keywords and their assigned priorities after the construction phase, following an iterative expert review process similar to established practices in the field~\cite{bross2013automatic,bloom2010unsupervised}, to ensure the correctness of the lexicon.

The use of both TF-IDF weighting scheme and the BoW for pre-processing the lexicon-based approaches is a well-established technique in information retrieval~\cite{manning2009introduction,salton1988term}. In the following, we describe the details of the TF-IDF used in our approach. 
\RestyleAlgo{ruled}
\begin{algorithm}[hbt!]
 \caption{Smart-Contract Vulnerability-related Keywords Lexicon Construction}
\label{algo1} 
\small

\SetKwInOut{Input}{Input}\SetKwInOut{Output}{Output}
\SetKwComment{comment}{\#}{}
\Input{(i) RG: review with code weaknesses from GitHub\
(ii) RC: review with code weaknesses from NVD\
(iii) KT: extracted key terms\
(iv) TFIDF: term frequency function used to create sorted words\
(v) Rj: a code review}
\Output{Lexicon with index numbers generated from TFIDF}
\BlankLine

\textbf{STEP 1:} Preprocess and concatenate RG and RC to create a corpus $C$\BlankLine
\textbf{STEP 2:} Extract KT from $C$ to create a list of key terms $T$\BlankLine
\textbf{STEP 3:}
\For{each $Rj$ in $C$}{
    \If{$Rj$ has been previously processed}{
        Continue 
    }
    \For{each term $Ti$ in $Rj$}{
        $F1 \gets TFIDF(term(Ti), Rj)$ 
        
    }
}
\textbf{STEP 4:} Save all the results in a file\BlankLine
\textbf{STEP 5:} Sort the words in $B$ based on their term frequency in descending order in F1\BlankLine
\textbf{STEP 6:} Select the top 250 terms and assign index numbers to the sorted words to create the lexicon\BlankLine
\textbf{STEP 7:} Return the lexicon
   
\end{algorithm}

\textbf{Details of the TF·IDF and BoW.} We use \textit{TF-IDF} function (i.e., Algorithm~\ref{algo1}-step 3) to apply the term-weighting algorithm~\cite{salton1988term} on the corpus of the code reviews. Let $C$ denote a corpus of code reviews and $Rj_c \in C$ denote a code review in the corpus. The term frequency, $\text{tf}(T_i, Rj_c)$, is defined as the number of times a term $T_i$ occurs in the code review $Rj_c$. Let $C_{T_i}$ denote the set of code reviews in the corpus $C$ that contain the term $T_i$. The number of code reviews in which the search term $T_i$ appears is denoted as $|C_{T_i}|$, while $|C|$ represents the total number of code reviews in the corpus $C$. The inverse document frequency (idf) of term $T_i$ in corpus $C$ is defined as $\text{idf}(T_i) = \log \frac{|C| + 1}{|C_{T_i}| + 1}$.Moreover, the addition of '+1' in the denominator, specifically '$|C| + 1$', prevents division by zero and addresses the scenario where a term occurs in every document in the corpus ($|C_{T_i}| = |C|$).

The TF-IDF weight of a term $T_i$ is proportional to the number of times it appears in a code review, $Rj_c$, and inversely proportional to the number of code reviews where the term occurs. In order to calculate TF-IDF, we used the following formula.

\begin{equation}\label{eq:tfidf}
\text{TF}\cdot\text{IDF}(T_i, Rj_c) = \text{tf}(T_i, Rj_c) \times \log\frac{|C| + 1}{|C_{T_i}| + 1}
\end{equation}

As shown in Formula~\ref{eq:tfidf}, TF-IDF is the combination of TF and IDF functions. When a term occurs in a document frequently, its TF-IDF weight increases, and when it occurs in a corpus frequently, it decreases. For instance, stop words are common terms that occur in a large fraction of textual descriptions of the code reviews and do not contribute much to discriminating them. \textit{Therefore, the idf part in TF-IDF helps to measure the importance of a term in the code review. If a term appears frequently in the code review, it may not be very helpful in identifying relevant reviews. The idf helps to downplay the significance of such terms.}
Moreover, we apply BoW function in the same way as TF-IDF. However, we use the following function. 
\begin{equation}
\text{BG}(T,Rj) = \sum_{i=1}^{n} [t_{i}=T]
\end{equation}
Where BG is the BoW function, $\sum_{i=1}^{n} [t_{i}=T]$ is the count of occurrences of the term $T$ in the code review $Rj$. It is important to note that BG function performs similarly to TF-IDF and represents the count of occurrence of $t_i$ in $R_j$.

\subsection{Model Construction}
We constructed a model for code review classification in smart contracts using a pre-trained DistilBERT model, a variant of BERT (Bidirectional Encoder Representations from Transformers), which has been distilled for faster and more efficient processing while retaining much of BERT's performance. The model architecture leverages the Transformer architecture, which has shown remarkable success in NLP tasks~\cite{sanh2019distilbert}.

The model consists of an input layer, a DistilBERT layer, a dropout layer with a dropout rate of 0.2 to prevent overfitting, followed by two dense layers with 128 and 64 hidden units, respectively. Dropout regularization is applied after the DistilBERT layer to prevent overfitting by randomly setting a fraction of input units to zero during training. The output layer, a fully connected dense layer with a softmax activation function, predicts the probability distribution over the classes.


\textbf{Input process layer}
In this layer, we utilize the HashingVectorizer~\cite{joulin2016bag} because it provides a computationally efficient method for preprocessing and transforming text data into fixed-size numerical vectors. It can effectively handle out-of-vocabulary words~\cite{weinberger2009feature}, as it does not require a pre-built vocabulary. This attribute makes it more robust when encountering new, unseen words that might not have been present in the training data. HashingVectorizer performs well with heterogeneous data~\cite{ganchev2008small}. However, HashingVectorizer may not capture semantic relationships between words as effectively as other sophisticated embedding techniques such as Word2Vec~\cite{mikolov2013efficient} or GloVe~\cite{pennington2014glove}. In this layer, we initialize the HashingVectorizer with a fixed number of features (10,000). It is then used to transform the training and validation texts into numerical vectors. By setting the number of features to 10,000, the HashingVectorizer creates a 10,000-dimensional vector representation for each input text.
The HashingVectorizer layer applies a hash function to the individual tokens (words) in the code review, mapping these hashed values to a fixed-size vector space. The vector representation is created by summing the hashed values for each token in the code review, resulting in a fixed-size vector representation for each code review, regardless of its original length. The hashing function can be represented by the following formula:
\begin{equation}
h(t) = t \bmod N
\end{equation}

where $h(t)$ is the hash function, $t$ is the input token, and $N$ is the number of features (in our case, 10,000). The hashing function maps the input token to a position in the vector space by taking the remainder of the division of $t$ by $N$.

\textbf{Output Layer (Dense, Softmax Activation):} The output layer consists of a dense layer with the number of units equal to the number of priority levels, which is 4 in our case. The softmax activation function is applied to the output layer for multi-class classification. This activation function converts the raw output scores from the neural network into probabilities, ensuring that the predicted probabilities for each class sum up to 1. The softmax function computes the probability $P(\text{class}_k)$ for each class $k$ using the following formulas:

\begin{equation}\label{eq:softmax1}
Z_i = \exp(a_i) \quad \text{for } i \in \{1, 2, \dots, K\}
\end{equation}

In Equation~\ref{eq:softmax1}, $a_i$ represents the raw output score for the $i$-th class produced by the neural network, and $K$ is the total number of classes ($K = 4$ in our case). The exponential function $\exp(\cdot)$ is applied to each $a_i$ to obtain positive scores $Z_i$ for every class.

\begin{equation}\label{eq:softmax2}
P(\text{class}_k) = \frac{Z_k}{\sum_{i=1}^K Z_i} \quad \text{for } i \in \{1, 2, \dots, K\}
\end{equation}

In Equation~\ref{eq:softmax2}, $P(\text{class}_k)$ represents the probability of the input belonging to the $k$-th class. To calculate this probability, the exponential score $Z_k$ for class $k$ is divided by the sum of the exponential scores for all classes. This normalization step ensures that the probabilities of all classes sum up to 1.

The model architecture includes a dropout layer with a dropout rate of 0.2 to prevent overfitting. This dropout layer is followed by two dense layers with 128 and 64 hidden units, respectively. Dropout regularization is applied after the DistilBERT layer to prevent overfitting by randomly setting a fraction of input units to zero during training. The output layer, a fully connected dense layer with a softmax activation function, predicts the probability distribution over the classes.

\subsection{Training Details}
In this work, we fine-tuned the hyperparameters of \praioritize to achieve optimal performance. We utilized the Sparse Categorical Crossentropy loss function, which is suitable for multi-class classification tasks. Additionally, we employed Sparse Categorical Accuracy as the evaluation metric to measure the model's performance.

The training process involved iterating over the dataset for a total of 10 epochs, with a batch size of 64 for each training iteration. Each epoch consisted of a training loop and an evaluation loop. During the training loop, the model parameters were updated using backpropagation to minimize the loss function. Dropout regularization with a dropout rate of 0.2 was applied as mentioned earlier.

During training process, we monitored the loss and accuracy metrics on both the training and validation datasets. At the end of each epoch, we evaluated the model's performance on the validation set using classification metrics. We also generated a confusion matrix to visualize the distribution of predicted labels compared to the ground truth labels. 
We employed Adam optimization algorithm with default settings for learning rate and other parameters. The descriptions of the reviews are first preprocessed using a HashingVectorizer with 10,000 features, followed by one-hot encoding of the labels. This prepares the data for input into \praioritize. 
We implement \praioritize using the open-source Keras library\footnote{https://github.com/keras-team/keras} built on top of TensorFlow~\cite{abadi2016tensorflow}. \praioritize is trained on the training dataset and evaluated on the validation dataset. The best model weights achieved by \praioritize are saved during training using a checkpoint callback based on validation accuracy. Finally, we assessed \praioritize performance by utilizing classification reviews and evaluation matrices.

\section{Experiment Evaluation}
\label{sec:results}

In this section, we describe our experiment and evaluation measures. Moreover, we present our findings.


    



\subsection{Baseline Selection}
\textbf{Baseline 1:} We utilize the approach proposed by Meng et al.~\cite{meng2022automatic} in which the text information of the code weakness along with the weakness explanation were used to classify code weakness reviews. The proposed approach utilizes BERT and TF-IDF to extract the features of the text information, then it trains machine learning classifiers (e.g. K-Nearest Neighbor) to classify the code weaknesses. We consider the textual features part and implement the model strictly as described in the paper, as the code is not provided in the paper. 


\textbf{Baseline 2:} We also considered DRONE as it is the most cited state-of-the-art approach by Tian et al.\cite{tian2013drone}. Drone proposed GRAY (ThresholdinG and Linear Regression to ClAssifY Imbalanced Data) to classify code weaknesses based on several features extracted from six dimensions, i.e., textual, temporal, author, related report, severity, and product. Because smart contracts are still in their early stages, most of these dimensions are not available in the code reviews yet, so we consider the textual dimension only. We implement the model precisely based on the description provided in the research paper, as the source code is not available.

\subsection{Evaluation Measures}

To evaluate the performance of \praioritize, we use commonly-used metrics in the literature, namely precision, recall, and F1-measure. 

Precision represents the ratio of correctly predicted instances to the total number of predictions. Recall, on the other hand, is the ratio of correctly predicted instances to the total number of actual instances. F1-measure is a harmonic mean of precision and recall, and it provides a balanced evaluation of the model's performance. In the following, we summarize the evaluation metrics used in our analysis:

\begin{equation}
\text{Accuracy: } \frac{\text{TP + TN}}{\text{TP + FP + TN + FN}}
\end{equation}

\begin{equation}\label{eq:measures}
\text{Precision: } \frac{\text{TP}}{\text{TP + FP}}
\end{equation}

\begin{equation}
\text{Recall: } \frac{\text{TP}}{\text{TP + FN}}
\end{equation}

\begin{equation}
\text{F1-measure: } 2 * \frac{\text{Precision * Recall}}{\text{Precision + Recall}}
\end{equation}

Where: 

\begin{itemize}
    \item True Positive (TP): The model correctly predicts a positive sample as positive.

\item True Negative (TN): The model correctly predicts a negative sample as negative.

\item  False Positive (FP): The model incorrectly predicts a negative sample as positive.

\item False Negative (FN): The model incorrectly predicts a positive sample as negative.
\end{itemize}

\section{Empirical Results}
\label{sec:empirical}
 In this section, we answer the proposed RQ and list our results and findings. 

\subsection{Zero-day Attacks in Smart Contracts}
Using the automatic analysis outlined in the methodology section~\ref{sec:framework}, we identified zero-day attacks in smart contracts. In our analysis, we excluded four CVEs (i.e., CVE-2020-17752, CVE-2018-17882, CVE-2018-18665, and CVE-2018-18666) due to unclear disclosure dates in third-party advisories, opting to omit them from the analysis.  

The analysis of vulnerabilities within smart contracts underscores the prevalent risk posed by zero-day attacks. The findings reveal that a substantial portion, approximately 77.22\%, of identified vulnerabilities were exploited by attackers before any mitigation measures could be implemented. 
This category represents vulnerabilities that were leveraged without prior disclosure or available fixes. Additionally, approximately 20.27\% of vulnerabilities were identified but not actively exploited, suggesting a potential window of opportunity for auditing teams to address these weaknesses before they are exploited. Conversely, a smaller fraction of vulnerabilities, accounting for 1.14\%, were patched by developers before any exploitation occurred. Nevertheless, a portion of vulnerabilities, approximately 1.37\%, remains of uncertain exploitation status as the dates were not clearly listed in the vulnerability information as mentioned earlier.  

Furthermore, we analyzed the sources of code reviews associated with each CVE. Our investigation revealed that these reviews primarily originate from three main sources: GitHub, OpenZeppelin forums, and Ethereum forums exclusively. Notably, GitHub emerged as the most prevalent source of code reviews across the CVEs analyzed. In contrast to traditional software, where CVEs may utilize specified prioritization systems, for instance, as seen in CVE-2018-13093 and CVE-2018-13095, the nature of smart contract vulnerabilities and their associated code reviews demonstrates a distinct reliance on community-driven platforms for vulnerability disclosure, prioritization, and patching.

\begin{tcolorbox}[colback=gray!10!white,colframe=gray!60!black,title=Summary] 
\begin{itemize}
    \item Among the analyzed smart contract vulnerabilities, 77.22\% were identified as zero-day attacks, indicating a significant prevalence of previously unknown vulnerabilities exploited by threat actors. 

\item Smart contract vulnerabilities lack dedicated tracking systems. Instead, the disclosure, prioritization, and patching of vulnerabilities rely heavily on community-driven platforms and associated code reviews.
\end{itemize}

\end{tcolorbox}
 
\subsection{Answers to RQ1:Prioritization of Smart Contract Code Weaknesses}
In order to evaluate the performance of the proposed automated labeling, an expert manually reviewed the labeled code reviews that were obtained by automatic labeling. In
the manual review, the expert removed all the labels generated by the automated labeling and labeled the data according to their priority level. The manual review is based on the terminology proposed in the Background section~\ref{sec:review}.  Then, we calculated the inter-rater agreement using the kappa coefficient~\cite{viera2005understanding} between the data labeled by the expert and the automatically generated labels. 
The resulting kappa coefficient is 0.92, which indicates a high level of agreement. We observed that most disagreement cases occurred within the critical and high-priority levels, revealing a tendency for ambiguity between these priorities in our analysis. This ambiguity suggests a lack of clear distinction in documentation between high and critical priorities. Additionally, our manual labeling process revealed inadequate descriptive terms for high-priority reviews, where terms such as 'weaknesses', 'issue' and 'vulnerability' were sometimes used interchangeably, leading to confusion. Furthermore, these reviews often lack sufficient information about the exploitability or potential impact of identified weaknesses. The distinction between high and critical-priority reviews lies in the severity of the vulnerabilities they expose and their potential for exploitation. Critical-priority reviews identify vulnerabilities with significant impact, allowing attackers unauthorized access to the smart contract and compromising its integrity and security. Conversely, high-priority reviews highlight weaknesses that, while impactful, may not immediately grant unauthorized access but still represent substantial weaknesses within the contract itself, affecting contract performance. However, these weaknesses are often poorly described in the reviews, contributing to underestimation of their impact and overlooking significant flaws. These differences extend beyond mere vulnerability identification, influencing both the impact and potential exploitability of identified weaknesses

Overall, our review shows that the proposed automated labeling performs similarly to the manual labeling in terms of inter-rater agreement. This finding supports the usefulness and validity of automatically prioritizing code reviews in smart contracts as a first step in the labeling process to reduce time and effort.

\begin{table}[]
\centering
\small
\begin{tabular}{@{}lllll@{}}
\toprule
\textbf{Priority Level} & \textbf{Precision} & \textbf{Recall} & \textbf{ F1-measure} & \textbf{Average  }  \\ \midrule
\textbf{Low}              &                   0.81 & 0.86 & 0.83 & 0.833                    \\
\textbf{Medium}           &                   0.81 & 0.83 & 0.82 & 0.82                     \\
\textbf{High}             &                  0.92 & 0.80 & 0.86   & 0.86                     \\
\textbf{Critical}         &                   0.95 & 0.99 & 0.97 & 0.97                     \\
\midrule
\end{tabular}
\caption{PrAIoritize performance}
\label{table:modelperformance}
\end{table}

As shown in Table~\ref{table:modelperformance}, PrAIoritize predicts the four priority levels with an average F-measure of 83\%, 82\%, 86\%, and 97\%, respectively. The F-measure for the critical level is the best and the medium priority level is the lowest. Low and medium levels are close in F-measure. Taking the mean of the F1-measures for all priority levels, regardless of their support, the macro average of F-measures is 87\%, indicating reasonably good performance in predicting priority levels. Nevertheless, we believe it is highly important for code review prioritization to have higher accuracy and F-measure for the critical level than other priority levels, which PrAIoritize is able to provide. 
Moreover, the consistency of F1-measure values across different priority levels highlights the reliability of the PrAIoritize model. The consistent performance across the four priority levels suggests the model's adaptability and effectiveness in handling various code review complexities and priorities.


\begin{table}[]
\centering
\small
\begin{tabular}{@{}llllll@{}}
\toprule
\textbf{Priority Level} & \textbf{Low} & \textbf{Medium} & \textbf{High} & \textbf{Critical} & \textbf{Average} \\ \midrule
\praioritize        &   0.83        & 0.82            &   0.86      &             0.97    & \textbf{0.87}\\
\cite{meng2022automatic}                  & 0.63              & 0.38                & 0.73              &   0.96                & \textbf{0.68}  \\

 DRONE                   & 0.29             & 0.10                & 0.32              &       0.47            &  \textbf{0.30}                  \\
BERT           &     0.75           &    0.80            &  0.82              & 0.97                  &     \textbf{0.83}             \\

T5                     &0.25            &  0.10               & 0.46              &  0.92                  & \textbf{0.43}                 \\

BiLSTM                     &0.79            &  0.77               & 0.66              &  0.95                    & \textbf{0.79}                 \\ 


RNN                     & 0.76                &  0.70                 &  0.52               &  0.95              & \textbf{0.73}                 \\ 
\bottomrule
\end{tabular}
\caption{Comparisons of F-measures of \praioritize versus other classifiers and baselines}
\label{tab:fmeasureall}
\end{table}

Moreover, we investigate how effective frequently used pre-trained models are as well as the selected baselines, namely DRONE and Meng et al.~\cite{meng2022automatic} for the same classification task. Table~\ref{tab:fmeasureall} shows the F-measures of \praioritize versus the two baselines and four popular pre-trained models for text classification tasks (i.e., BERT, T5, BiLSTM, and RNN). We utilized identical parameters and configurations as those detailed in the PrAIoritize method section~\ref{sec:framework}. We also briefly describe the selected models in the Background section~\ref{sec:review}. We notice that Meng et~al.~\cite{meng2022automatic} can predict the four priority levels with F-measures equal to 0.63, 0.38,0.73, and 0.96 from Low to Critical levels, which means it can assign critical priority levels correctly while it faces challenges in correctly identifying Medium priority levels. This may be attributed to the proposed feature extraction method (i.e., BERT) or the KNN classifier by Meng et~al.~\cite{meng2022automatic}. 

DRONE's F-measure results show that it is struggling in assigning correct Low, Medium, and High levels to the code reviews. While it performs better with the Critical level (i.e., F-measure equals 0.47), it remains the lowest-performing model when compared to other models in our experiment. We believe it is because of the linear regression of the Gray classifier, which is not well-suited to our dataset. 

Comparing the two baselines with the result of \praioritize, we observe that we can improve the F-measures for the four priority levels, with the Medium level slightly less than the rest of the priority levels. Considering the overall average F-measures, \praioritize   outperforms Meng et~al.~\cite{meng2022automatic} baseline by 27.94\% and DRONE baseline by approximately 194.9\%. However, we only consider the textual dimension of DRONE implementation. Therefore, in cases where the dataset includes the other five dimensions (i.e., temporal, author, related report, severity, and product as proposed in DRONE paper), DRONE may surpass our model's performance.

Moreover, we studied the performance of BERT, T5, BiLSTM, and RNN in comparison to \praioritize performance. Among these well-known models, we notice that RNN has the poorest average F-measure. PrAIoritize's average F-measure outperforms BERT by approximately 4.82\%, T5 by approximately 102.33\%, BiLSTM by approximately 10.13\%, and RNN by approximately 19.18\%


\begin{table}[]
\centering
\small
\label{F1measures}
\begin{tabular}{@{}llllll@{}}
\toprule
\textbf{Priority Level} & \textbf{Low} & \textbf{Medium} & \textbf{High} & \textbf{Critical} & \textbf{Average} \\ \midrule
\praioritize                 &   0.86           & 0.83                &  0.80               & 0.99                  &          \textbf{0.87}         \\
\cite{meng2022automatic}              & 0.64             & 0.43                &   0.67            & 0.96                  &  \textbf{0.68    }            \\

DRONE                   &   0.37           & 0.08               &    0.42           &    0.34               &   \textbf{0.30}               \\
BERT          &      0.85        & 0.81                 & 0.73               &  0.97                  &     \textbf{0.84}             \\

T5                     &0.25            &  0.10               & 0.46              &  0.92                  & \textbf{0.43}                 \\ 
BiLSTM                   &  0.86           &  0.84               &     0.58          & 0.94                  &             \textbf{0.80}     \\


RNN                    &   0.72           &   0.93              &   0.37            &   0.95                &           \textbf{0.74}       \\

\bottomrule

\end{tabular}
\caption{Comparisons of Recall measures of \praioritize versus other classifiers and baselines}
\label{tab:recall}
\end{table}
\begin{table}[]
\centering
\small
 \begin{tabular}{@{}llllll@{}}
\toprule
\textbf{Priority Level} & \textbf{Low} & \textbf{Medium} & \textbf{High} & \textbf{Critical} & \textbf{Average} \\ \midrule
\praioritize  &   0.81             & 0.81                & 0.92                &  0.95                    &  \textbf{0.87  }           \\

\cite{meng2022automatic}                 & 0.64              & 0.50                & 0.61              &   0.96                & \textbf{0.68}                  \\
 
DRONE                   & 0.24             &   0.12              & 0.25              &  0.73                 & \textbf{0.34}                 \\
BERT          & 0.67              &    0.79              & 0.95              & 0.99                  &    \textbf{0.85}              \\

T5                     &0.25            &  0.10               & 0.46              &  0.92                  & \textbf{0.43}                 \\ 
BiLSTM                   &   0.74           &  0.72                & 0.76               &  0.97                   & \textbf{0.80}                 \\


RNN                      &  0.81           &  0.56               &      0.84    &  0.96            &     \textbf{0.79}            \\

\bottomrule
\end{tabular}
\caption{Comparisons of Precision measures of \praioritize versus other classifiers and baselines}
\label{tab:precision}
\end{table}
Table~\ref{tab:recall} shows the recall measures for all baselines and models versus \praioritize recall measures for the four priority levels. \praioritize outperforms Meng et al.~\cite{meng2022automatic} baseline with 27.94\% in terms of the recall measure and DRONE by approximately 190.0\%. It also further improves the recall of BERT by approximately 
3.57\%, T5 by approximately 
102.33\%, BiLSTM by approximately 
8.75\%, and RNN by approximately 
17.57\%.
Finally, Table~\ref{tab:precision} presents the precision measure of \praioritize and the rest models. Similar to F-measure and recall, \praioritize significantly improves the precision of Meng et al.~\cite{meng2022automatic} by 27.94\%, achieves a higher precision of DRONE by 155.88\%, and the selected pre-trained models (i.e., BERT, T5, BiLSTM, and RNN) by 2.35\%, 
102.33\%, 
8.75\%, and 
10.13\% respectively. 

\begin{figure}[ht] 
\includegraphics[width=0.4\textwidth]{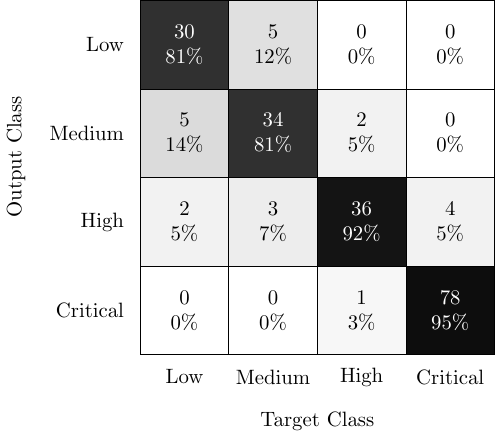}
       \caption{Confusion matrix for the classification results of the \praioritize model.}
       \label{fig:Confusion1}
\end{figure}
Figure~\ref{fig:Confusion1} shows a detailed breakdown of the classification results produced by the \praioritize model, as depicted in the confusion matrix. It illustrates the distribution of predicted labels compared to the ground truth labels. Each row in the matrix represents the actual (i.e., ground truth) class, while each column corresponds to the predicted class.
For low-priority weaknesses, the model achieves a noteworthy accuracy, correctly classifying 30 instances with minimal misclassifications. Similarly, in identifying critical-priority weaknesses, the model excels, accurately classifying 78 instances with very few misclassifications. However, for medium and high-priority weaknesses, the model's performance is less consistent. While it correctly identifies a majority of medium-priority weaknesses (i.e., 34 instances), there are a notable number of misclassifications, particularly as low and high priority. Similarly, in the case of high-priority weaknesses, the model shows decent accuracy, but there are noticeable misclassifications across other priority levels. These results suggest that while the model demonstrates satisfactory performance in code weakness prioritization, there is room for improvement, particularly in reducing misclassifications across different priority levels to enhance overall effectiveness.

\begin{tcolorbox}[colback=gray!10!white,colframe=gray!60!black,title=Summary] 
\begin{itemize}
    \item \praioritize effectively prioritizes code weaknesses in smart contract code reviews, outperforming both baseline models and state-of-the-art pretrained models.
\end{itemize}
\end{tcolorbox}

\section{Discussion and Implications}
\label{sec:implications}
Our results demonstrate that smart contract code weaknesses can be effectively prioritized automatically during the code review process. Our proposed automated prioritization approach, \praioritize, leverages DistilBERT and shows superior F-measure, precision, and recall compared to traditional models such as RNN and BiLSTM.
This disparity suggests that BERT-based models possess a better understanding of the semantics embedded within code reviews, enabling them to identify nuanced weaknesses more effectively.

DistilBERT and BERT, by leveraging transformer-based architectures, excel in capturing contextual information and semantic relationships within the text. This capability allows them to grasp the intricate details and nuances present in code reviews, leading to more accurate prioritization of weaknesses. Their performance highlights the importance of utilizing advanced natural language processing (NLP) techniques for code review prioritization tasks, especially in fields where precise comprehension of textual data is paramount.

Conversely, models such as RNN and BiLSTM, although performing reasonably well, do not exhibit the same level of semantic understanding as BERT and DistilBERT. Their reliance on sequential processing may limit their ability to capture long-range dependencies and intricate semantic nuances present in code reviews. As a result, while they achieve respectable precision levels, they may not fully exploit the contextual information embedded within the code reviews, leading to slightly inferior performance compared to BERT-based models.

Additionally, automated labeling for a corpus with limited keywords is useful as a first step to reduce time and effort, allowing for initial categorization of code reviews. However, predictive models, such as those leveraging DistilBERT, offer additional benefits by providing a more comprehensive understanding of the data and capturing nuances that automated labeling may overlook. Furthermore, predictive models can adapt to evolving datasets and provide insights that automated labeling alone may not capture, enhancing the robustness and accuracy of the overall process.
Furthermore, when comparing DistilBERT with T5, an intriguing observation arises. T5 differs significantly in its approach. T5 operates under the paradigm of text-to-text transfer learning, where it is trained to generate target text sequences from source text sequences in our case the code reviews. This approach allows T5 to learn complex mappings between input and output text, enabling it to effectively capture semantic relationships and contextual information within code reviews. However, compared to our approach (i.e., employing DistilBERT), T5  requires more computational resources due to its architecture, potentially leading to longer inference times. However, it is possible that increasing the number of epochs during training could improve T5's performance.

In summary, the results underscore the significance of leveraging advanced NLP techniques, particularly transformer-based models such as DistilBERT, for code weakness prioritization task in smart contracts. These models reveal superior precision by effectively understanding the semantics embedded within code reviews. While traditional models like RNN and BiLSTM perform adequately, they fall short of the nuanced understanding exhibited by transformer-based models. Additionally, T5 presents an alternative approach to capture semantic relationships within code reviews. However, its effectiveness might be influenced by its heightened resource demands and the need for extensive training.

\subsection{Implications}

\textbf{Implications for Software Practitioners:}
For software practitioners, \praioritize offers a valuable approach to enhancing smart contract development and auditing processes. By leveraging PrAIoritize's capabilities, practitioners can  prioritize code reviews more effectively, ensuring that critical code weaknesses are addressed promptly. The automated triage provided by \praioritize accelerates the identification of critical code weaknesses in smart contracts, minimizing the risk of overlooking urgent code weaknesses. Additionally, \praioritize facilitates early prediction of critical code weaknesses, enabling developers to promptly recognize and address potential risks. Third-party auditors can benefit from PrAIoritize's automated prioritizing capabilities, focusing their efforts on high-impact areas and providing valuable insights to project stakeholders. Furthermore, \praioritize improves the patch management process by automatically categorizing weaknesses based on priority, reducing development time and costs.

\textbf{Implications for Researchers:}
Researchers can leverage \praioritize to advance understanding and exploration in the field of smart contract security. By analyzing the dynamic evolution of code weaknesses and their corresponding priorities, researchers can uncover patterns that reveal changes in smart contract security landscape. PrAIoritize's predictive and prioritization capabilities can be used to gain insights into review practices and code weakness management in smart contract field. Additionally, \praioritize offers researchers a means to prioritize code weaknesses, facilitating subsequent studies on the correlation between developer attributes, code review methodologies, and code weakness prioritization. Future research can focus on developing multi-objective approaches to address the conflicting tasks of early code review prediction and prioritization effectively. Overall, \praioritize provides researchers with a powerful  means to investigate code weakness prioritization in smart contracts and correlate them with socio-technical aspects of code review, thereby enhancing our understanding of smart contract security.

\textbf{Implications for Tool Builders:}
\praioritize offers tool builders significant opportunities to develop advanced automation tools tailored specifically to smart contract development and auditing needs. By integrating \praioritize into automated bots, tool builders can create automated solutions that assist developers in predicting and prioritizing code review requests. These bots can be smoothly integrated into existing development ecosystems, such as Gerrit~\footnote{https://www.gerritcodereview.com/}, to improve the review process by monitoring newly submitted code reviews and prioritizing them. Moreover, \praioritize can significantly benefit project maintainers tasked with managing large codebases featuring numerous pending code reviews, such as smart contracts and dependent Dapps. By leveraging PrAIoritize's prioritization capabilities, maintainers can efficiently manage and prioritize code reviews, even in projects with hundreds of code reviews and changes. For example, the tool can provide maintainers with a sorted list of critical weaknesses, enabling them to allocate resources efficiently for prompt mitigation. Additionally, tool builders can empower users to personalize \praioritize priority levels according to their review preferences, such as setting thresholds for allocated review effort or adjusting prioritization criteria. This customization ensures that \praioritize adapts to the specific needs of individual developers or project teams, enhancing its usability and effectiveness in diverse development environments. Overall, by leveraging PrAIoritize's capabilities, tool builders can develop innovative solutions that improve the code review process and enhance overall development efficiency in smart contract ecosystems.

These implications can significantly enhance the smart contract code review process, providing valuable assistance in efficiently identifying and prioritizing code weaknesses. By leveraging the capabilities of \praioritize, developers and auditors can optimize their review workflows, ensuring that critical code weaknesses are addressed promptly and effectively. Furthermore, the suggested research implications offer opportunities for gaining insightful insights into how developer behavior and review practices impact code weakness management.

\section{Related Work}
\label{sec:related_work}

This section describes related research on code review and code weakness priority prediction.

\subsection{Prioritizing code reviews in Software Engineering}

An early work by Greiler et al.~\cite{greiler2016code} highlighted the significant challenge project maintainers face in prioritizing code review requests. This is due to the manual inspection required to determine which reviews to handle first. Several studies have attempted to address this challenge by proposing models to prioritize code reviews based on their likelihood of being addressed (~\cite{fan2018early}, ~\cite{islam2022early}). Another study by Zhao et al.~\cite{zhao2019improving} introduced a learning-to-rank approach to prioritize code changes. Moreover, Fan et al.~\cite{fan2018early} proposed a Random Forest (RF)-based approach to predict whether a code change would be addressed or abandoned. Subsequently, Islam et al.~\cite{islam2022early} enhanced the approach proposed by Fan et al.~\cite{fan2018early} with PredCR, utilizing Light Gradient Boosting Machines (LGBM) and a reduced feature set. 

However, while these studies have addressed code review prioritization using multiple features, we propose prioritizing code reviews based on the semantics of the code weaknesses within the reviews and their impacts (i.e., textual features). Therefore, we employ natural language processing and large language models. Furthermore, given our focus on prioritizing smart contract code reviews with code weaknesses, we classify the reviews into four urgency-based categories for patching.

\subsection{Code weakness priority prediction in Software Engineering}
Several methods have been suggested in the literature for improving the quality of software. These approaches can be grouped into duplication detection and classification, code weakness triage, and code weakness localization. Many studies have focused on predicting the priority of code weaknesses, and some of these studies are based on deep learning. For example, Fang et al.~\cite{fang2021effective} proposed a method that uses graph convolutional networks and a weighted loss function for the prediction of code weakness-patching priority. Another study by Li et al.~\cite{li2022tale} used deep multitask learning to develop an approach called PRIMA, which simultaneously learns both the code weakness category prediction task and the priority prediction task. 
Other studies are based on machine learning, for instance, Tian et al.~\cite{tian2015automated} proposed a machine learning model for priority prediction based on features extracted from six dimensions: temporal, textual, author, related report, severity, and product. Valvida-Garcia et al.~\cite{valdivia2014characterizing,valdivia2014characterizing} used the experience features of reporters to create blocking code weakness prediction models based on various classical machine learning classifiers. 
Zhou et al.~\cite{zhouexploratory} conducted a study to examine the impact of source code file feature sets on the accuracy of code weakness priority classification. The results of the study demonstrated that source code file feature sets did not perform as well as textual description features in code weakness classification. Tran et al.~\cite{tran2020analysis} compared different machine learning methods for evaluating the severity and priority of software code weaknesses. They suggested using an approach based on optimal decision trees to assess the severity and priority of new code weaknesses. 
Another study by Haung et al.~\cite{huang2022bug} developed a model for predicting multi-class code weakness priority that integrates sentiment and community-oriented sociotechnical features from both users and developers. The model was validated in various scenarios, including within-project and cross-project. The results of the study indicate that including assignee and reporter features from sociotechnical perspectives can improve prediction performance.

\subsection{Code reviews  prioritization and  prediction in Smart Contracts}
As far as we are aware, there has been no prior attempt to automatically predict the priority of code reviews and code weaknesses in smart contracts. Research into the prioritization of code reviews and weaknesses within smart contracts is still in its infancy. Some studies have attempted to define severity levels of smart contract code weaknesses  (i.e., \cite{zhang2020framework,chen2020defining}). 
However, the severity levels defined in these studies are limited to a few well-known vulnerabilities in smart contracts. The code reviews indicate that vulnerabilities in smart contracts evolve over time, just as smart contracts themselves do, leading to the emergence of new code weaknesses and resulting in new vulnerabilities. Moreover, severity levels in smart contracts are not yet standardized, similar to the priorities assigned to code weaknesses. It is important to note that severity is determined by customers, while priority is determined by developers \cite{tian2015automated}. 

\section{Threats to Validity}
\label{sec:threats}
A potential threat to internal validity arises from considering code weakness types based only on keywords to assign priority levels, potentially resulting in inconsistencies within the same code weakness type, as priority levels may vary within it. To mitigate this, we assessed the impact of each vulnerability as listed in the code review. Additionally, an expert reviewed the assigned priorities to ensure their accuracy. An additional threat to internal validity, while we have assigned priority levels to smart contract weaknesses in our study, it is important to notice that these levels do not follow any standardized classifications due to the lack of standardization in smart contracts~\cite{soud2022fly}. To mitigate this threat, our priority levels were identified based on keywords extracted from CVE records, related literature, and the reported issues on GitHub. Additionally, as the Ethereum platform and the language of smart contracts evolve, the prioritization of weaknesses may change to accommodate new advancements and features. Therefore, prioritization levels provided in this study should be considered as a starting point and may require adjustment based on future developments in the field. Additionally, future studies could enhance the methodology by incorporating severity levels (i.e., when available) into the labeling process. Another potential challenge to internal validity can be our exclusive focus on Solidity code weaknesses. This limited focus might not encompass all the code weaknesses in all smart contract languages, affecting the generalizability of our findings. It is important to be careful when applying our results to other languages, as they can differ significantly in code weaknesses and behaviors.
Another potential threat to internal validity arises from the significant similarity among reported vulnerabilities in the dataset. This high level of similarity poses a unique challenge in accurately distinguishing the priority of each code weakness. To mitigate this threat, we employed techniques such as exploring the context of code weaknesses in the contract and applying DistilBERT to understand the semantics of the weaknesses. However, it is essential to interpret the results cautiously, considering the inherent complexities of the dataset. 

One potential threat to external validity with our proposed approach is its generalizability. To address this, we trained and evaluated \praioritize on a dataset comprising two reliable sources to further verify the effectiveness of our approach and reduce the risk of threats to external validity. A potential threat to the construct validity of our study is the choice of evaluation metrics. However, the metrics we selected (precision, recall, and F-measure) are widely used in the literature and have been adopted in numerous previous studies, including the selected baselines~\cite{tian2013drone,meng2022automatic}. Another concern with the construct validity of our study is the distribution of the four levels of priority across the two data sources. While this may potentially impact the performance of \praioritize to some extent, the high performance across data sources with four priority levels suggests the effectiveness of our approach.

\section{Conclusion}
\label{sec:conclusion}

Smart contract vulnerabilities and weaknesses are widespread. To address these weaknesses, auditors and engineers conduct manual code reviews on platforms such as GitHub. This study aims to enhance the security and quality of smart contracts by automatically predicting the priority of smart contract code weaknesses during the code review process. We present \praioritize, an approach that leverages NLP and LLMs for prioritizing code weaknesses in smart contracts. Evaluation of \praioritize demonstrates its effectiveness in accurately prioritizing code reviews. Our approach provides a reliable means of automatically predicting code weakness priorities for smart contracts, highlighting the utility of NLP and multi-class LLMs in review prioritization. Future work will involve expanding our research by including code reviews from additional OSS projects and exploring a wider array of pre-trained models commonly used for similar tasks.






\bibliographystyle{model1-num-names}
\bibliography{mybibfile.bib}

\end{document}